\documentclass[usegraphicx,usenatbib]{mn2e}

\usepackage{journalcodes}

\newcommand{\sub}[1]{_{\rmn {#1}}}
\newcommand{\OmegaP}{\Omega\sub{P}}
\newcommand{\CMHOG}{\textsc{cmhog2}}
\newcommand{\HII}{H\,\textsc{ii}}

\begin{document}

\title[Constraining corotation from spiral shocks]
{Constraining corotation from shocks in tightly-wound spiral galaxies}
\author[D.~M.\ Gittins and C.~J.\ Clarke]
{D.~M.~Gittins$^1$\thanks{Email: dgittins@ast.cam.ac.uk}
and C.~J.~Clarke$^1$\thanks{Email: cclarke@ast.cam.ac.uk}\\
$^1$Institute of Astronomy, Madingley Road, Cambridge CB3 0HA}

\maketitle

\begin{abstract}
We present a new method for estimating the corotation radius in tightly wound spiral galaxies,
through analysis of the radial variation of the offset between arms traced by the
potential (P-arms) and those traced by dust (D-arms).
We have verified the predictions
of semi-analytical theory through hydrodynamical simulations and have examined the uniqueness
of the galactic parameters that can be deduced by this method.
We find that if the range of angular offsets measured at different
radii in a galaxy is greater than around $\pi/4$, it is possible
to locate the radius of corotation to within $\sim 25$\%.
We argue that the relative
location of the P- and D-arms provides more robust constraints on the galactic
parameters than can be inferred from regions of enhanced star formation (SF-arms), since interpretation
of the latter involves uncertainties due to reddening and the assumed star formation law.
We thus stress the importance of K-band studies of spiral galaxies.
\end{abstract}


\section{Introduction}

Theories of the structure, origin, development and effects of spiral arms in galaxies
have occupied researchers ever since such structure was discovered in 1845.
Many galaxies are observed with tightly-wound,
regular spiral arms covering a significant fraction of their disc.
The commonest model of this phenomenon
is based on the Lin-Shu hypothesis of spiral structure \citep*{lin69}, which 
states that quasi-steady spiral patterns exist in the stellar disc,
rotate at a pattern speed $\OmegaP$, and persist for many orbits.
The orbits of the stars are organised into `kinematic density waves', which
increase the stellar density in a spiral pattern.
The stars themselves
drift through the pattern inside the corotation radius, as they orbit with
an angular velocity $\Omega > \OmegaP$.

The spiral density wave picture is by no means the only theory of spiral structure,
but provides a very successful description.  The possible origins,
driving mechanisms and long-term support of such density waves has been
the subject of a great deal of literature, and this subject is beyond the scope of this
paper; see reviews by \citet{kaplan74} and \citet{toomre77}.
It is generally accepted that spiral waves, if they do not naturally arise
through instabilities, may be driven by nearby companions
or central bars, and that under the right conditions they may persist for
many rotations, and it is from this starting point that we proceed in this paper.

Galaxies that are optically classified as spirals usually have a clear
spiral structure with one dominating mode, and arms are usually logarithmic
in shape (e.g.\ they have a constant pitch angle) \citep{kennicutt81,kennicutt82,garciagomez93,ma02}.
However, galaxies
not classified as spirals are now known commonly to show spiral structure
in their stellar discs.  The old stellar population, which will trace the
mass of the disc, can be observed in longer wavelengths such as the K or K$^\prime$ bands.
\citet{schweizer76} found `broad spiral patterns' in the discs of six spiral
galaxies, concluding that smooth spiral structure exists in their mass distributions.
\citet{elmegreen84} examined several galaxies of different types, suggesting
a class of galaxy which has flocculent structure at blue wavelengths and
a spiral structure in K.
More recently, \citet{block94} showed that optical classification does not
constrain the structure of stellar discs, finding smooth spiral structure
in galactic discs independently of their optical structure.
Many galaxies optically classified as flocculent, when observed in K$^\prime$,
show regular spiral structure \citep{thornley96,grosbol98,seigar03}.
Most of these have a central bar, and some also have a neighbour \citep*{seigar03}.
The presence of regular spiral structure, but flocculent structure in bluer
wavelengths, has also arisen in numerical simulations \citep{berman02}.
We make the distinction between three separate arm components: the
peaks of the underlying stellar mass distribution, approximately tracing the perturbations
in the gravitational Potential (the P-arms), the shock fronts of the gas,
which will generally be observed from Dust (the D-arms), and the ridges of
enhanced Star Formation (the SF-arms).

Many tightly-wound spiral galaxies are known.  Assuming that the model
of a rotating spiral potential is accurate, the parameters of that potential
are fundamental to the galaxies themselves.  An understanding of the structure
of spiral arms is important in considering the processes affecting the ISM
flowing through them.  In this paper, we focus on the determination of the
corotation radius.

We discuss the formation of single-shock flows in the gaseous
disc, and show how the angular offset between these shocks (the D-arms) and the minima
of the gravitational potential (the P-arms) could be used to constrain corotation.  In section
\ref{model}, we introduce the model spiral potential, to which the gas responds.
Section \ref{response} investigates the resulting gas flow using a well known semi-analytical
approach, and compares the results to those obtained by numerical methods.
In section \ref{offset}, different components of spiral arms are discussed, and the
offset function $\Theta$ is introduced.  Various ways to constrain corotation
are summarised in section \ref{corotation}, and the use of the offset function to provide
an additional constraint is explained.  Finally, our findings are summarised
in section \ref{conclusions}.


\section{Model spiral potential}
\label{model}

The rigidly rotating potential of a spiral galaxy can
be decomposed into a static, axisymmetric part and
a spiral perturbation.  We define the gravitational potential $V$
(in galactocentric polar coordinates $R,\theta$) as
\[
	V(R,\theta,t) = V\sub{R}(R) + V\sub{S}(R,\theta,t)
\]
where $V\sub{R}$ is the axisymmetric potential and $V\sub{S}$ is the spiral perturbation.

The axisymmetric part represents the unperturbed velocity curve.
In this paper, the following velocity curve is used:
\[
	v(R) = v\sub{max} \sqrt{ F\sub{b}\epsilon\sub{b}R \exp(-\epsilon\sub{b}R)
	+ 1 - \exp(-\epsilon\sub{d}R) }
\]
where $v\sub{max}$ is the limit of the circular velocity at large radius,
$\epsilon\sub{d}$ and $\epsilon\sub{b}$ are the
inverse disc and bulge scale lengths (respectively) and $F\sub{b}$ is the
bulge strength parameter \citep{contopoulos86}.
The potential $V\sub{R}$ that produces this velocity curve is
\[
	V\sub{R}(R) = v\sub{max}^2 \left( \ln R + E_1(\epsilon\sub{d} R) - F\sub{b}
		e^{-\epsilon\sub{b} R} \right)
\]
where the exponential integral $E_1$ is defined as
\[
	E_1(x) = \int_x^\infty{\frac{e^{-u} \; {\rm d}u}{u}}.
\]

A spiral perturbation with one mode is given by the general form
\[
	V\sub{S} = A(R,t) \cos(\chi)
\]
where the spiral phase $\chi$ is defined as
\[
	\chi = \Phi(R) - m (\theta - \OmegaP t)
\]
and the pattern has $m$ arms.  The pattern rotates rigidly
at angular velocity $\OmegaP$.  The radial shape of the arms is
set by the function $\Phi(R)$, and at constant radius, the potential
is sinusoidal in $\theta$ with period $2\pi/m$.
The overall amplitude is set by $A(R,t)$.  The equipotentials
of each mode are spirals, which have an inclination $i(R)$ to circles.
Trailing logarithmic arms, with a constant inclination $i$, correspond
the choice
\[
	\Phi(R) = -\frac{m}{\tan i} \ln ( R )
\]

It is convenient to
define the perturbation strength $F(R)$ as the ratio of the amplitude of the perturbation
force to the axisymmetric force:
\[
	F = \frac{\left| \bmath{\nabla} V\sub{S} \right|\sub{max}}
		{\left| \bmath{\nabla} V\sub{R} \right|}
	= \frac{m A}{v^2 \sin i}
\]
where the circular velocity at radius $R$ is $v$.
The radial and time variation of the amplitude can be chosen arbitrarily.
A suitable radial choice is to specify \citep{contopoulos86}
\[
	A(R,t) = A_0 R \exp(-\epsilon\sub{s} R)
\]
where $\epsilon\sub{s}$ is the inverse spiral scale length, and the overall
amplitude at any time is set by $A_0$.
In this paper we are interested in steady solutions in which the potential
has reached a stable state, so that $A$ is independent of time.
$A_0$ can most conveniently be set by requiring that the maximum relative
spiral strength takes a specified value $F\sub{0}$ at some radius $R_0$.

\begin{figure}
\includegraphics[width=0.45\textwidth]{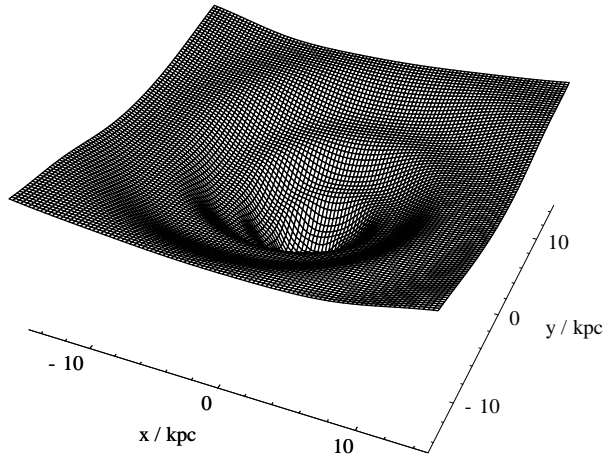}
\caption
{Potential surface with parameters $F\sub{b}=0$, $\epsilon\sub{d}^{-1}=1.5$ kpc,
$\epsilon\sub{s}^{-1}=10$ kpc, $\sin i=0.2$, $m=2$ and $F\sub{0}=.6$ at $R_0=8.5$ kpc.
A very high value of $F$ is chosen to produce a spiral perturbation visible on this
figure.}
\label{pot_surface}
\end{figure}

An example potential surface, with an exaggerated spiral perturbation, is shown in figure
\ref{pot_surface}.

\section{Response of gaseous disc}
\label{response}

A relatively small perturbation can have a significant effect on the
flow of gas through it.  Large deviations from the circular flow can
lead to the formation of shocks.  In this section we examine the
formation of single-shock solutions in the flow of isothermal gas through
the spiral potential.
Since the intent is to examine the large-scale behaviour,
and ignore small-scale processes, approximations of initial uniformity and
isothermality are appropriate.  Here, the sound speed used is an `effective' sound speed
equal to the velocity dispersion of the ISM, a good approximation for the time and length
scales involved \citep{cowie80}.

\subsection{Semi-analytical technique}

In this paper we repeat the procedure used in
\citet*{shu73} for finding non-linear solutions (containing a shock) to the
asymptotic equations of gas flow under a spiral perturbation.
The derivation of the equations and descriptions of the procedure are
given in \citet{roberts69} and \citet{shu72} and are
only briefly described here.

\begin{figure}
\includegraphics[width=0.45\textwidth]{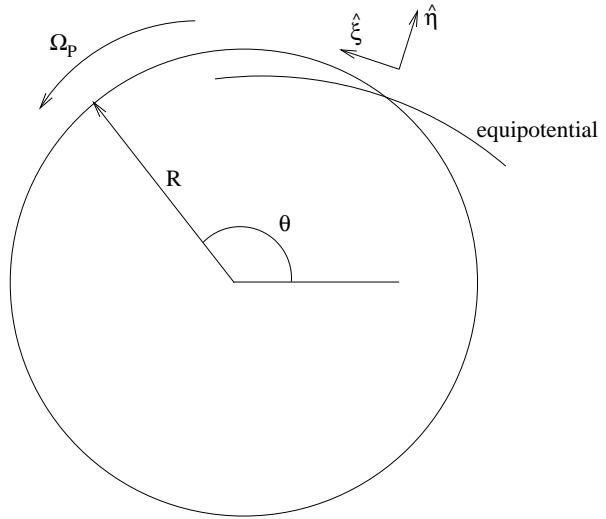}
\caption
{The coordinate system used in this paper.}
\label{diagram_coords}
\end{figure}

Solutions are expressed in  `curvilinear' coordinates $(\eta,\xi)$ that vary perpendicularly
to and parallel to the equipotentials, respectively (see figure \ref{diagram_coords}).
These are defined in
a stationary frame of the spiral perturbation.  They are defined by\footnote{
These coordinates differ slightly to those in \citet{shu72} and \citet{roberts69}.
}
\[
	d\eta = -k dR + m d\theta
\]
\[
	d\xi = -m \frac{dR}{R} - kR d\theta
\]
where
\[
	k = \frac{\partial \Phi(R)}{\partial R} = \frac{-m}{R \tan i}
\]
where the last result is for the case of logarithmic arms.
A circle at constant $R$ will therefore pass through $m$ periods of $\eta$.
Specifically we define
\begin{equation}
	\eta = -\chi + \pi
\label{def_eta_general}
\end{equation}
so that $\eta=0$ always corresponds to the potential minimum.  The passage
from one arm to the next corresponds to a period of $2\pi$ in $\eta$, regardless
of the number of arms.

The velocity of the gas is written in components $(u\sub{\eta},u\sub{\xi})$ in
this coordinate system.  These are then expressed as the sum of the `base'
flow $(u\sub{\eta 0},u\sub{\xi 0})$ (the equilibrium circular flow when the perturbation is not present,
in the stationary frame of the perturbation) and
a perturbed velocity $(u\sub{\eta 1},u\sub{\xi 1})$.
In the specific case of logarithmic arms the coordinates can be defined as follows:
\[
	\eta = \frac{m}{\tan i} \ln( \epsilon\sub{s} R) + m(\theta - \OmegaP t) + \pi
\]
\[
	\xi = -m \ln (\epsilon\sub{s} R) + m \tan i (\theta - \OmegaP t).
\]

The differential equations in $u\sub{\eta}$ and $u\sub{\xi}$, and the technique for
locating appropriate solutions, are given in appendix \ref{appendix}.
The boundary condition for any solution is that it be periodic in $\eta$ with period $2\pi$.
Solutions are found along streamlines under the approximation that
they are at nearly constant radius, and the surface density $\sigma$ is then found
from the velocity.
The solution at a given radius is controlled by seven parameters, under the
approximation that they are constant along a streamline:

\begin{description}
\item $m$ (the number of arms)
\item $\OmegaP$ (the pattern speed of the arms)
\item $i$ (the local pitch angle)
\item $F$ (the local relative arm strength)
\item $v$ (the unperturbed circular velocity)
\item $\kappa$ (the local epicyclic frequency)
\item $a$ (the effective sound speed of the gas)
\end{description}

Base-supersonic flow is defined as any flow with $u_{\eta 0}>a$, so that
in the stationary frame of the arms, the unperturbed flow perpendicular to the
equipotentials is supersonic.  Similary, base-subsonic flow has $u_{\eta 0}<a$.
In general, if the flow in a galaxy is mostly base-supersonic, there
will be a region (the base-subsonic region) in which it is not.  This,
of course, occurs either side of the corotation radius in a band where
the relative velocity $R(\Omega - \OmegaP)$ is sufficiently small.
In this region, the response of the gas, and the asymptotic flow
solutions, will be qualitatively different.
The base-subsonic region is bounded by the radii satisfying
$R \sin i (\Omega - \OmegaP) = \pm a$.

If for the whole solution $u_\eta<a$ (entirely subsonic) or $u_\eta>a$
(entirely supersonic), the solution will not contain a shock.  This will occur under conditions
where the sound speed $a$ is very high or low, or the strength $F$ of the perturbation
is sufficiently small.
In the limit of small $F$, the solutions reach the first-order response, which is
sinusoidal in $u$ and $v$ (and hence $\sigma$), such that $u_\eta \propto -\cos \eta$ and
$u_\xi \propto \sin \eta$.  The maximum density (and minimum of $u_\eta$) occur at the centre
of the potential well, i.e.\ at $\eta=0$.

If the strength $F$ of the perturbation is gradually increased, starting from an
entirely subsonic solution, then at some point the flow will be sufficiently perturbed
that $u_\eta$ will pass the sound speed $a$.  Any solution must now contain a sonic point,
at which the gas accelerates to supersonic velocities, and a shock.  Such solutions are no
longer symmetric.  A consistent solution consists of a closed curve in the $(u_\xi,u_\eta)$ plane
containing a single isothermal shock, with a periodicity in $\eta$ of $2 \pi$.  Such
a solution only exists for certain choices of parameters.

\begin{figure}
\includegraphics[width=0.45\textwidth]{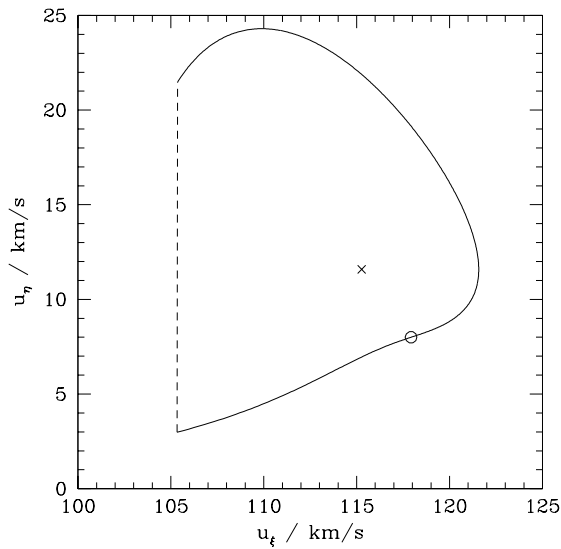}
\caption
{Velocity curve in $(u_\xi,u_\eta)$ plane of the solution at $R_0=8$ kpc and $F=0.05$.
The cross marks the unperturbed velocity.  The sonic point is marked by a circle.  The shock
is represented by the dashed line.}
\label{vu_shock}
\end{figure}

\begin{figure}
\includegraphics[width=0.45\textwidth]{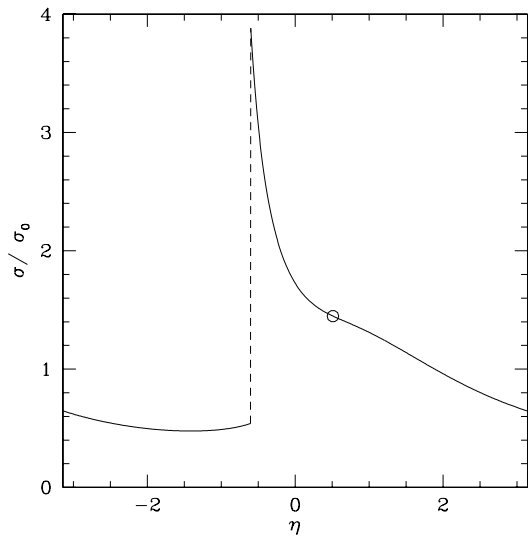}
\caption
{Surface density profile $\sigma(\eta)$ corresponding to the solution shown in figure \ref{vu_shock}.
The sonic point is marked by a circle.}
\label{rho_shock}
\end{figure}

Figure \ref{vu_shock} shows an example velocity curve
in $(u_\xi,u_\eta)$ for a solution in a tightly wound spiral, and
the corresponding density profile through the shock is shown in figure \ref{rho_shock}.

\subsubsection{Existence of a solution}

It is not always possible to find solutions containing a shock for any choice of parameters.
The equations are based on the approximation that the gravitational field is in the
$\eta$ direction (and hence that the shock is parallel to the equipotentials), which
will hold for large $|k|R$ or (equivalently) $\sin i \ll 1$, so that the spiral
arms are tightly wound and $\Phi(R)$ varies rapidly.  In more open spirals the
equations therefore break down.  Also, the parameters
are approximated as constant along a streamline, which will be a good approximation
only if $F$ is not much larger than $\sin i$.  In our experience, solutions
are difficult or impossible to find for $F \ga 0.1$ and $\sin i \ga 0.2$
(with all other parameters as indicated in table \ref{standardmodel}).

Within parameters that satisfy these approximations, solutions still may not
exist.  In particular, solutions in the base-subsonic region are difficult to find,
as discussed in appendix \ref{appendix}.
In many cases, although shock solutions
can be found, no solution of the correct periodicity exists.

Many solutions show secondary peaks in the density.  These occur especially
near the ultraharmonic resonances, where higher harmonic terms in the solution
become important.  If a secondary peak crosses the sonic line, a second shock
will occur.  It will therefore become impossible to locate a single-shock solution,
since a singularity will be present at the second sonic point.
It is possible to extend the method described here to include a
secondary shock \citep{shu73}, but this has not been attempted in this paper.

\subsubsection{Results}

\begin{figure}
\includegraphics[width=0.45\textwidth]{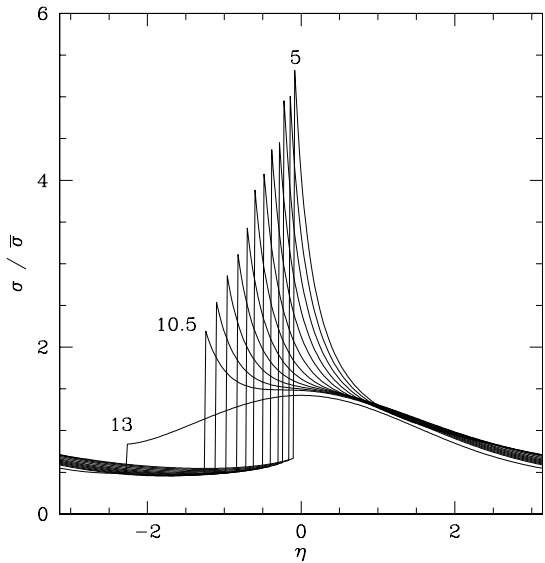}
\caption
{Density profiles $\sigma(\eta)$ for the standard model, at radii from $R=5$ kpc to
$R=10.5$ kpc (indicated) in steps of $0.5$ kpc.  The profile at $R=13$ kpc is also shown.}
\label{profiles}
\end{figure}

A `standard model' is defined, with parameter values as given in table \ref{standardmodel}.
Density profiles for the standard model are shown in figure \ref{profiles} for a range
of radii.  No profiles are shown for radii between 11 and 12.5 kpc, as no solutions exist.
The figure shows the smooth transition toward the base-subsonic density profiles, in which
the broad density peak at $\eta=0$ appears and the shock is small.
This figure also indicates that the shock location moves to smaller values of $\eta$
with increasing radius.  At small radii, the shock occurs almost at the potential minimum,
but as the radius increases and the shock becomes weaker, it moves upstream toward
the potential maximum.

\begin{table}
\caption{Parameter values of the standard model.}
\begin{tabular}{rr@{\,}l}
$F\sub{b}$		&	$0$	&				\\
$\epsilon\sub{d}^{-1}$	&	$1.5$	& kpc				\\
$\epsilon\sub{s}^{-1}$	&	$10$	& kpc				\\
$m$			&	$2$	&				\\
$\sin i$		&	$0.1$	&				\\
$\OmegaP$		&	$13$	& km s$^{-1}$ kpc$^{-1}$	\\
$a$			&	$8$	& km s$^{-1}$			\\
$F(R_0) \equiv F_0$	&	$0.05$	&				\\
$v(R_0)\equiv v_0$	&	$220$	& km s$^{-1}$			\\
$R_0$			&	$8.5$	& kpc				\\
\end{tabular}
\label{standardmodel}
\end{table}

\subsection{Numerical calculations}

The response of an isothermal gas disc can be computed with numerical
hydrodynamical codes.  In this section, results obtained using two-dimensional
Smoothed Particle Hydrodynamics (2D SPH) and a two-dimensional
Piecewise Parabolic Method (PPM) code are compared with the results
of the semi-analytical analysis.  Numerical codes are not restricted
by the approximations of the semi-analytical approach, nor to results containing
a single shock, and should therefore be able to capture more detailed structure.

\subsubsection{Two-dimensional Smoothed Particle Hydrodynamics}

\begin{figure}
\includegraphics[width=0.45\textwidth]{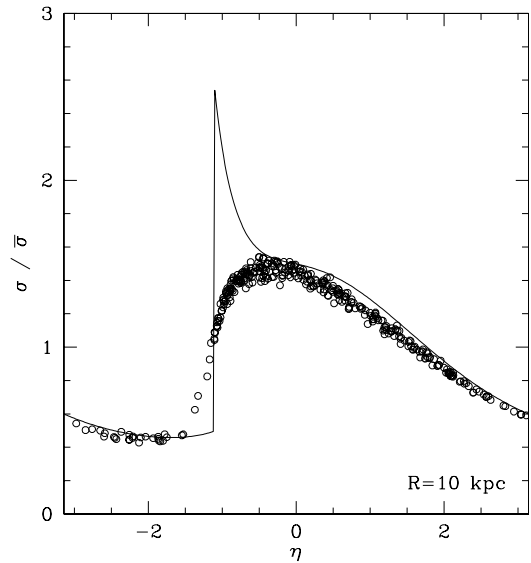}
\caption
{Density profile $\sigma(\eta)$ of the standard model at $R=10$ kpc as computed
by the semi-analytical method (solid line), compared to results extracted from
a 2D SPH calculation (circles).}
\label{SPHpr_MW1m2_10}
\end{figure}

Calculations with 2D SPH showed a good general agreement to the results
of the semi-analytical analysis, but did not resolve the sharp shocks
accurately.  A tendency for the SPH particles to `clump' together unphysically
was overcome by using a triquintic spline smoothing kernel,
but this results in decreased spatial resolution  \citep[see][for details]{thesis}.  This problem may be
alleviated by adjusting the time dependence of the smoothing lengths in
the way described in \citet{englmaier97}.  A typical example of results
obtained with 2D SPH is shown in figure \ref{SPHpr_MW1m2_10}.

\subsubsection{Two-dimensional Piecewise Parabolic Method}

A 2D PPM code called \CMHOG\ was provided by P.~Teuben.  The code uses PPM
in the Lagrangian remap formalism, and includes isothermal hydrodynamics
in polar coordinates.  The code is based on that described in \citet*{piner95}; for
a description of the Piecewise Parabolic Method, see \citet{colella84}.

\begin{figure}
\includegraphics[width=0.45\textwidth]{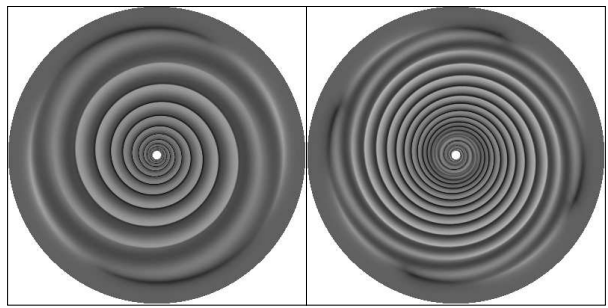}
\caption
{Surface density maps from calculations of the standard model (left) and
with $m=4$ (right) at $t=1$ Gyr.  The maximum radius is 20 kpc.}
\label{CMm2m4_pics}
\end{figure}

Results from calculations of the standard model and a model with $m=4$ are shown
in figure \ref{CMm2m4_pics}.  Sharp shocks
were formed as expected, and the strength of the shocks fades towards the corotation radius.
Results from further calculations, with variations
of the strength $F$ and the pattern speed $\OmegaP$, were very similar.

\begin{figure}
\includegraphics[width=0.45\textwidth]{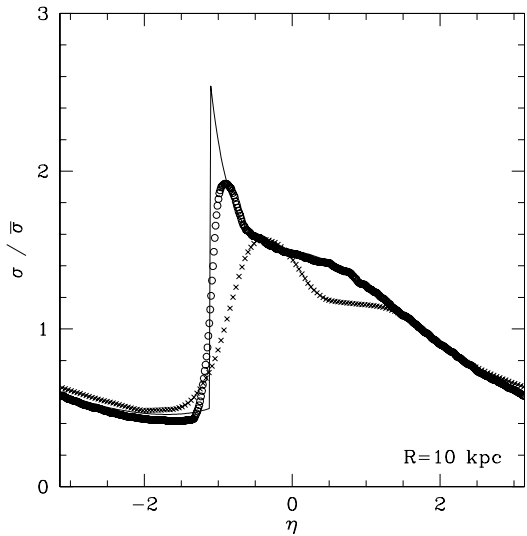}
\caption
{Density profile $\sigma(\eta)$ of the standard model at $R=10$ kpc as computed
by the semi-analytical method (solid line), compared to results from
the grid-code calculation (circles).  Also shown are the results from the
same calculation with $1/4$ the number of radial zones (crosses).}
\label{CMpr_MW1m2_10}
\end{figure}

\begin{figure}
\includegraphics[width=0.45\textwidth]{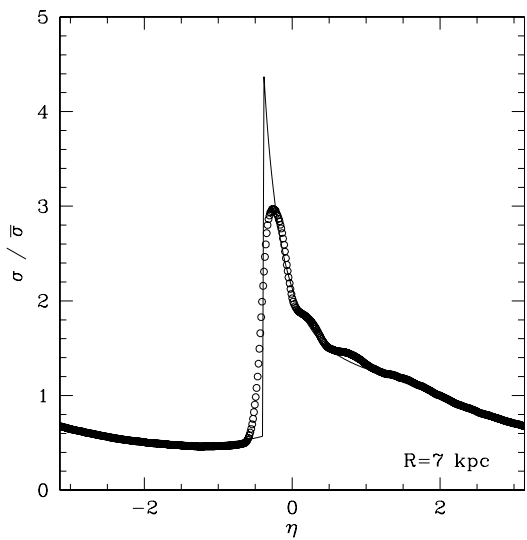}
\caption
{Comparison as in figure \ref{CMpr_MW1m2_10}, at $R=7$ kpc.}
\label{CMpr_MW1m2_7}
\end{figure}

\begin{figure}
\includegraphics[width=0.45\textwidth]{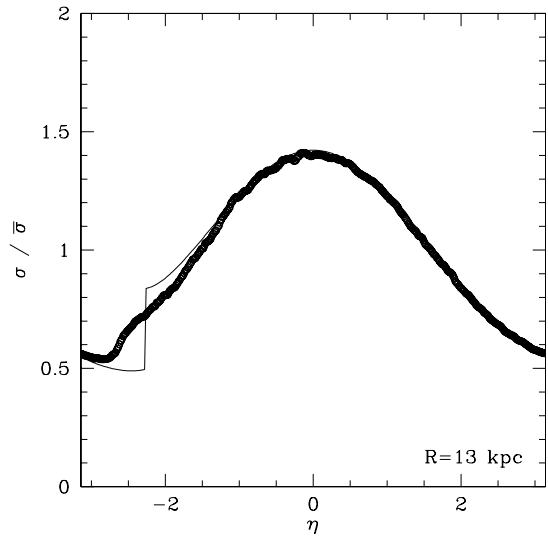}
\caption
{Comparison as in figure \ref{CMpr_MW1m2_10}, at $R=13$ kpc.  This
radius is in the base-subsonic region, and is close to the 6:1 resonance.}
\label{CMpr_MW1m2_13}
\end{figure}

Results from the standard model are compared to semi-analytical results in
figures \ref{CMpr_MW1m2_10}, \ref{CMpr_MW1m2_7} and \ref{CMpr_MW1m2_13}, at radii
$R=10$, 7 and 13 kpc, respectively.  Recall that a period of $2\pi$ in $\eta$
always corresponds to the passage from one arm to the next, regardless of the number
of arms.
The agreement is excellent in all three cases,
and the narrow density peaks are reproduced by the grid code.  The sharp shock front
is spread over several zones, but it is reproduced much more accurately
than in the SPH code.  All the density profiles show some small oscillations
in the density downstream of the shock.  At $R=13$ kpc, there is once again a discrepancy between
the semi-analytical prediction and the calculation result, probably due to the proximity
of the 6:1 ultraharmonic resonance.

The agreement with the semi-analytical profiles improves as the resolution
of the grid-code is increased.  Figure \ref{CMpr_MW1m2_10} shows
the result obtained with 150 radial zones, compared to the full 600, for comparison.
This trend held up to the limits of available resolution,
and presumably, with more computing resources, ever more accurate
agreement could be achieved.

\begin{figure}
\includegraphics[width=0.45\textwidth]{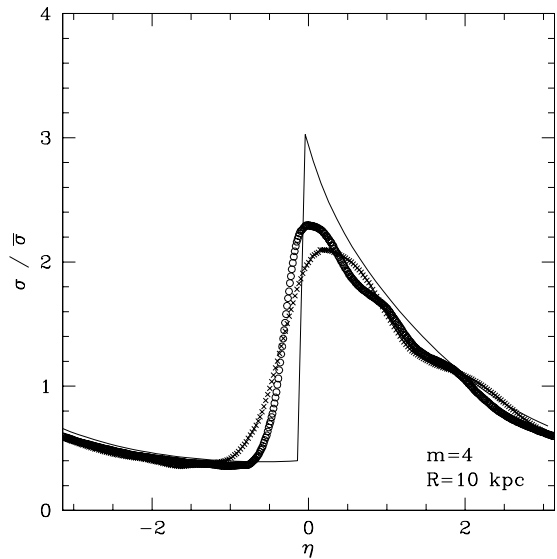}
\caption
{Comparison as in figure \ref{CMpr_MW1m2_10} with $m=4$.  The result with
half the angular resolution is also shown (crosses).}
\label{CMpr_MW1m4_10}
\end{figure}

Figure \ref{CMpr_MW1m4_10} shows a comparison at 10 kpc with the calculation
using $m=4$.  The general shape of the profile is again well-represented, but
in this case the position of the shock is slightly further upstream in the grid-code
result than the semi-analytical prediction.  This is a general result when $m=4$.
The reason for the difference is not clear.  The number of grid zones
covering one arm is half that in the $m=2$ calculations, so doubling the
resolution might remove the offset.  Such high resolution is impractical
at time of writing.  However, when the angular resolution is halved instead,
the predicted shock location is not obviously changed, which suggests that the effect
is not resolution dependent.

\begin{figure}
\includegraphics[width=0.45\textwidth]{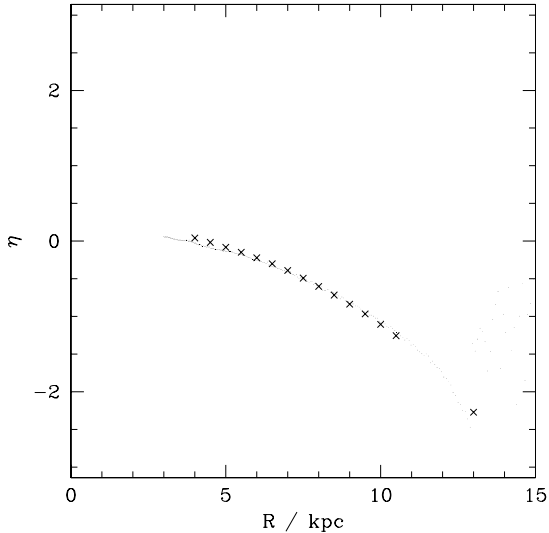}
\caption
{Shock location vs.\ radius in the standard model, as measured from the grid-code
calculation (points) and predicted from the semi-analytical results (crosses).}
\label{CMoffsets_m2}
\end{figure}

\begin{figure}
\includegraphics[width=0.45\textwidth]{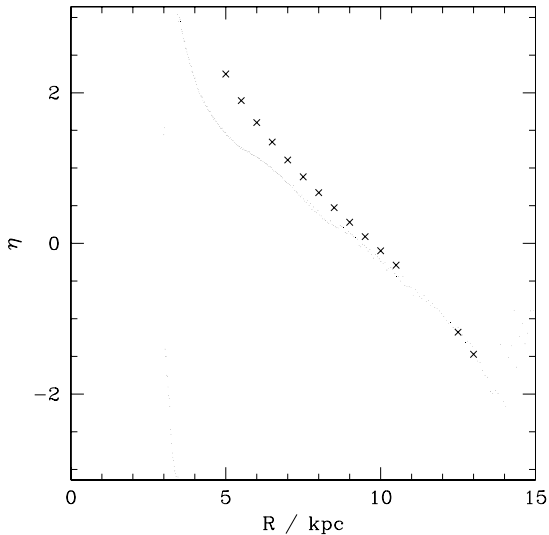}
\caption
{Shock location vs.\ radius in the $m=4$ model, as measured from the grid-code
calculation (points) and predicted from the semi-analytical results (crosses).}
\label{CMoffsets_m4}
\end{figure}

The shock location in the results
corresponds very closely to the predicted location from the semi-analytical
results in all cases where $m=2$, at radii between the inner Lindblad resonance
and the start of the base-subsonic region.
Figure \ref{CMoffsets_m2} plots the measured location of the shock as a function
of radius, from the calculation of the standard model.  The positions predicted
from the semi-analytical theory are also shown, where they can be found.  In figure
\ref{CMoffsets_m4}, the results for the model with $m=4$ are given.  When $m=2$,
there is excellent agreement between the two.  At large radii, in the base-subsonic region,
the shocks become so weak that the shock location in the grid-code results
cannot be determined accurately.
The comparison with $m=4$ shows
that the grid-code consistently places the shock further upstream than the
prediction of the semi-analytical theory, as discussed above.

Similar comparisons were made from calculations using $F=0.03$ and
$\OmegaP = 19.5$ km s$^{-1}$, and in both cases, very close agreement was seen.
Numerical results confirm the predictions of the semi-analytical method
in parameter regimes where such results are possible.  Grid-code calculations were also
carried out for parameters where no semi-analytical solutions exist,
such as those with larger inclinations (i.e.\ more open arms) or lower sound speeds.
In general, under these conditions, the gas failed to settle to a steady solution,
and a much more complex structure was seen involving multiple shocks and
significant secondary peaks in the density.  These more complicated cases
are left for further investigation.

\begin{figure}
\includegraphics[width=0.45\textwidth]{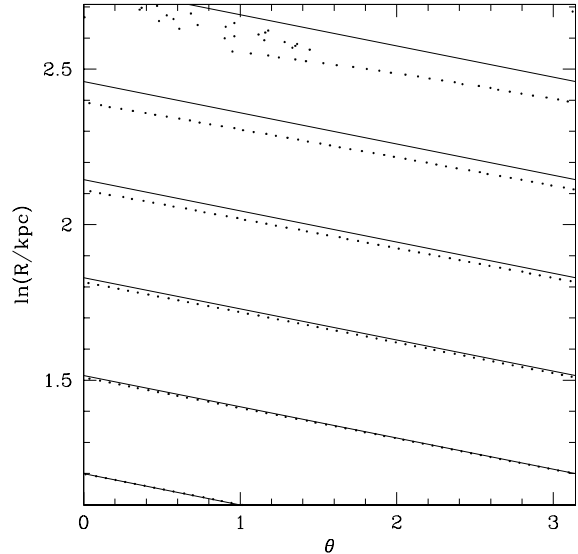}
\caption
{Shock location with $m=2$ measured from the grid-code calculation (points)
over the range 5--15 kpc.  Potential minima are marked as lines.  Only the range
$0<\theta<\pi$ is shown since the grid-code only calculates half the plane.}
\label{CMoffsets_m2_logspace}
\end{figure}

\begin{figure}
\includegraphics[width=0.45\textwidth]{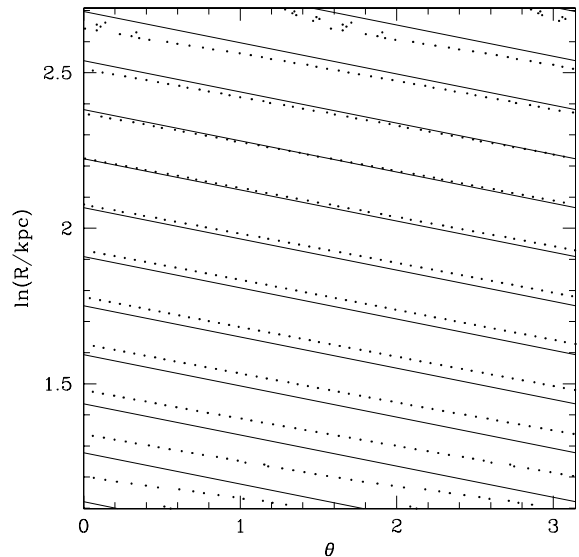}
\caption
{Plot as in figure \ref{CMoffsets_m2_logspace}, with $m=4$.}
\label{CMoffsets_m4_logspace}
\end{figure}

In figure \ref{CMoffsets_m2_logspace}, the shock location with $m=2$ is plotted as
$\ln R$ vs.\ $\theta$
and compared to the potential minima, which are of course straight lines on such
a plot.  The shocks are near the minima at small radii, and move toward the maxima
as corotation is approached, until they become too small for the grid-code to locate and the points
scatter.  The result with $m=4$ is shown in figure \ref{CMoffsets_m4_logspace}.  In this
case, the shocks move from the maxima at small radii to cross the minima, and continue
to move to the next maxima.  The shocks thus complete a transition from one arm to the
next.

\section{Spiral arm components and the offset function}
\label{offset}

\subsection{Shock locations}

The `offset function' referred to in this paper, $\Theta(R)$, is defined as
\[
	\Theta = m(\theta\sub{shock} - \theta\sub{min})
\]
where the potential minimum lies at angle $\theta\sub{min}$ and the shock is
at angle $\theta\sub{shock}$.  (The offset is defined to lie in the range $-\pi < \Theta \leq \pi$).
For tightly wound spirals, in which the streamlines
are almost circular, this is almost equivalent to $\eta\sub{shock} - \eta\sub{min}$.
Since $\eta$ is defined in such a way that the minima of the potential correspond
to $\eta=0$ (equation \ref{def_eta_general}), $\eta\sub{min} \equiv 0$, and
so
\[
	\Theta = \eta\sub{shock}.
\]
This very convenient result allows the offset
to be read directly from density profiles $\sigma(\eta)$, whether generated
semi-analytically or in numerical calculations.

The value of $\Theta(R)$ is difficult to predict without performing a full calculation.
It will be expected to vary with the relative velocity of the gas and the potential,
as it is the deceleration and acceleration of the gas that causes a shock to form
in the first place.  The offset should, therefore, vary systematically with radius
in a galaxy, since the relative velocity $R(\Omega - \OmegaP)$ decreases with increasing
radius towards corotation.

The existence and variation of this offset have been noted before.
\citet{roberts69} noted, from calculations of semi-analytical solutions, that
`the shock lies just on the inner side of the background [i.e.\ potential] spiral
arm', indicating that $\Theta$ was generally small and negative.
\citet{shu72} report a value of $\Theta=-72^\circ$ for their calculations
of the flow in the solar neighbourhood.  \citet{yuan81} generated synthetic
colour profiles across a spiral arm, based on semi-analytical solutions of the flow.
They state that `the shock \ldots\ always occurs on the inner side of the potential minimum'
($\Theta < 0$), and for their calculations adopt a constant value $\Theta=-30^\circ$
(in their paper $\Theta$ is given the symbol $\Delta_0$).  This result, that
$\Theta$ should be small and negative, is often tacitly assumed to be correct
in discussions of spiral arms.  For example, \citet{seigar98} state that
`the fact that dust lanes [associated with shocks] appear on the trailing
[i.e.\ upstream, inside corotation]
edges of [potential] arms is itself evidence for the large-scale shock scenario'.
The systematic variation of $\Theta$ with radius, however, does not
seem to have been discussed in the literature.

\citet{elmegreen93} performed particle-based simulations of a galaxy,
in which gas clouds and stars were represented by interacting particles.  In
the somewhat open ($\sin i \simeq 0.34$) spiral arms that formed in the disc,
they found that `the location
of the [gas] shock front changes from the inside of the [stellar] arm inside corotation to the outside
of the arm outside corotation', that is, $\Theta < 0$ inside corotation and $\Theta > 0$ outside.
This general trend is in agreement with the usual predictions discussed above.
However, the shock front is very poorly resolved in their results, and a detailed
examination of $\Theta$ would not be possible.

\citet*{kikuchi97} discuss angular separations between different arm components arising
in an unstable galactic disc. 
Their analysis differs from ours in that they constrain all components to share a common velocity
field, thus negating the possibility of shocks in the gas phase.
They nevertheless find that the ordering of the gas and stellar arms reverses at corotation, and suggest the utility of this
result in constraining the location of corotation in observed spirals.

\subsection{Observational characteristics of spiral arms}
\label{threearms}

In discussing the observational characteristics of a model spiral galaxy,
it is very important to be precise about what is meant by an `arm'.
Since a spiral arm will affect different components of the disc in different
ways, arms traced by one phenomenon may appear very different to arms
traced by another.

In general, one expects to see {\em three} spiral arm patterns.
The first is that traced by the potential minima, or the maximum
surface density.
In fact, these two definitions will not coincide exactly, since the potential
is always more curved in $R$ than the underlying density.  The effect is that
the potential minima are slightly more tightly wound than the peaks of
the surface density, but we will disregard this difference here.
These arms, created by the mass variations in the old stellar
population, are expected to have a smoothly varying structure.
The second
set of arms is traced by the shock front of the gas (where it exists).  In strong
shocks, this will also represent the maximum gas density.  Since dust is generally
taken to be a tracer of shocked gas, these arms will also be traced by
dust lanes.  The third set of arms is the region of enhanced star formation.  The existence
of this third arm, linked to the other two, relies on the assumption that the
shocked gas promotes star formation in the ISM.  There will then be some
finite `onset' time, during which the early stages of star formation take place.
This is followed by 
an increased density of young stars, \HII\ regions and so on, which will trace the
region of star formation.  This third set of arms would, therefore, be expected to
occur downstream of the second.
These three arms are labelled, in this paper, the P-arm (for Potential), the D-arm
(for Dust) and the SF-arm (for Star Formation).

With this definition, $\Theta$ refers to the offset between the P-arm and the D-arm,
irrespective of the existence or location of any SF-arm.
If $\Theta(R)$ is a constant,
then the D-arms will be identical to the P-arms, but rotated by an angle $\Theta/m$, and the two
will therefore have identical pitch angles.  If, however, $\Theta$ becomes more negative with increasing
radius (e.g.\ the shock moves further upstream of the potential minimum at larger radii),
then the D-arm should be more tightly wound than the P-arm.

The relative pitch angles of the D-arm and SF-arm are also likely to be different.
If the assumption of a constant {\em time} offset from the D-arm to the SF-arm (associated
with the onset of star formation) is made, then the angular offset will be proportional
to the relative angular velocity of the gas $\Omega - \OmegaP$.  Since this decreases
to zero as radii approach corotation, the offset will also decrease to zero.  The SF-arm would
therefore be expected to be more tightly wound than the D-arm.

\begin{figure}
\includegraphics[width=0.45\textwidth]{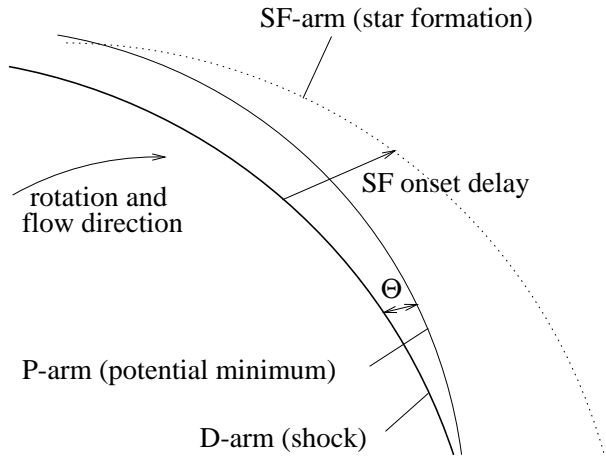}
\caption
{Diagram of the three types of spiral arm and their separations.  In this diagram
$\Theta$ becomes more negative with radius, so that the D-arm is more tightly wound
than the P-arm.  The star formation time delay is constant, but since the relative
velocity of the gas to the P-arm is smaller at larger radii, the angular delay is smaller.
Consequently the SF-arm is more tightly wound than the D-arm.}
\label{armpitch_diagram}
\end{figure}

Figure \ref{armpitch_diagram} is a diagram of the three arms, in the particular
case that $\Theta$ becomes more negative with radius and the star formation onset time is constant.
Under both of these circumstances, the pitch angles of the arms are expected to
follow $i\sub{P} > i\sub{D} > i\sub{SF}$.  \citet{visser80a} makes the point that
a constant $\Theta$ is required for the P-arms and D-arms to have the same pitch angle,
and comments that in M81 the tendency is for both the `gas and dust' (D-arms) and `\HII\ regions
and young stars' (SF-arms) to be tighter than the potential arms.

The three types of spiral arm are observed in different ways.  Observing the P-arm
requires observations at longer wavelengths, that isolate the older stellar population in the
disc.  These arms would probably, therefore, be observed as peaks in the K or I bands.
The D-arms, defined by the locations of shocks, should be traced by dust lanes
or radio emission ridges.  If these arms correspond to the peaks of the gas density,
then they should also be traced by peaks of CO emission (and other tracers of dense
molecular gas).  Finally, the SF-arms can be traced by bluer light, such as the B band,
or by H$\alpha$ emission from \HII\ regions.  Determining the location and extent of the SF-arms
is hampered by the effects of dust reddening.

Generally, the strongest CO emission is seen to coincide with the dust
lanes and nonthermal ridges, in M51 for example \citep{lo87}, and all of these are displaced from the
peak of H$\alpha$ emission \citep{tilanus88,nakai94}.  The offset from CO peaks
to \HII\ regions is seen in most grand design spirals, such as M31 \citep{ichikawa85}.
More recent observations have detected
a sequence of $^{12}$CO to $^{13}$CO to H$\alpha$ across an arm in M51 \citep{tosaki02},
supporting the idea of a gradual collapse of molecular clouds to higher densities in the onset of
star formation.  One exception is the observation of \citet{lord91} that
peaks of CO emission are offset from the dust lanes in an arm of M83, which they
attribute to dense clouds penetrating through the shock layer while a more diffuse
component is compressed to a shock.

In principle, therefore, the positions of the D-arm and P-arm can be observed.
This would then allow a direct measurement of $\Theta(R)$.
The resulting data
will enable constraints to be placed on the parameters of the galaxy's spiral
structure, as discussed in the next section.


\section{Locating the corotation radius in galaxies}
\label{corotation}

The determination of the corotation radius (or, equivalently, the pattern speed)
in spiral galaxies has been attempted by various methods for various galaxies.  The
assumption that a pattern speed exists at all implies that a rigidly rotating
spiral potential is present.

The simplest way to estimate the corotation radius is to associate it with the maximum extent
of observed spiral arms.  The idea that star formation is promoted by the shock compression
means that at corotation (where there can be no shock compression), star formation
should be absent.  The radius of the outermost \HII\ region has therefore been used
as an estimate of the corotation radius \citep[e.g.][in M81]{rots75}.  This method
is not reliable, since longer exposures tend to reveal further features at larger
radii in galaxies, and \HII\ regions may exist outside corotation.  An alternative, but similar,
approach is to associate the inner edge of observed spiral features with the
inner Lindblad resonance, on the
basis that spiral waves are not expected to extend beyond the Lindblad resonances.
However, several investigations have discussed the possibility of spiral patterns
extending within the inner Lindblad resonance \citep{elmegreen98,englmaier00,martini03},
so this method is unlikely to be accurate without detailed modelling.
More sophisticated analyses associate various observational features with a range of resonances
in a galaxy to constrain the corotation radius \citep*{elmegreen90,elmegreen92}.

Alternatively, if the velocity curve (and preferably other parameters such as velocity dispersions
and mass models) of a galaxy are known, then a corresponding spiral structure can
be simulated, based on a specified pattern speed and the spiral density wave theory.
The pattern speed can then be adjusted until the best fit is achieved.  This approach
was taken by \citet{visser80b} in an analysis of M81, and more recently for several
galaxies \citep*{kranz03}.

Another approach makes use of the predicted radial variations in the arm amplitude.
These are associated with interference between outward and inward travelling spiral
waves, and fitting the oscillations can locate corotation \citep*{elmegreen89}.
The `modal' approach \citep{bertin89} involves examining the full range
of normal modes of the galactic disc and their relative amplitudes.  The technique has been
applied to M81 \citep{lowe94}.  Other methods include:
matching radial oscillations in the velocity curve to predicted velocity fields
\citep{yuan69a}; the `geometric phase' method
\citep{canzian97}, which associates corotation with a shift from singly to triply
symmetric structure in the velocity field; and the integrated continuity equation
method \citep{westpfahl98}, in which asymmetries in the observed disc are exploited
to constrain $\OmegaP$.

Of course, values determined by different methods rarely agree exactly;
table \ref{M81table} lists pattern speeds derived for the spiral pattern in M81
from a variety of sources in the literature.  This indicates a spread of some 25\% in one
of the most studied spiral galaxies.

\begin{table}
\caption{Some pattern speeds of M81 reported in the literature.}
\begin{tabular}{r@{}ll}
\hline
\multicolumn{2}{c}{$\OmegaP\ /$ km s$^{-1}$ kpc$^{-1}$}	&	Source	\\
\hline
17&.4		&	\citet*{shu71}		\\
26&.4		&	\citet*{roberts75}	\\
20&		&	\citet{rots75}		\\
18&.3		&	\citet{gottesman75}	\\
18&		&	\citet{visser80a}	\\
20&		&	\citet{berman82}	\\
17&\ $\pm 2$	&	\citet{sakhibov87}	\\
26&		&	\citet{elmegreen89}	\\
24&		&	\citet{lowe94}		\\
23&.4 $\pm 2.3$	&	\citet{westpfahl98}	\\
\hline
\end{tabular}
\label{M81table}
\end{table}

A commonly suggested method of locating corotation arises from examination
of the separate arm components.  If star formation occurs at some time offset after the
passage of a spiral shock, then the relative location of the D-arm and the SF-arm
should be reversed inside and outside corotation.  The radius at which these
arms cross should, therefore, correspond to corotation.  This point is mentioned
by \citet{grosbol98}, but they failed to observe such a crossing.

\citet{dixon71} proposed that observations in the B and V bands could allow the variation
of stellar age across an arm to be determined, and that corotation should correspond
to the radius at which this trend reverses angular direction.
\citet{puerari97} attempted to locate corotation in two galaxies by this method,
performing a Fourier analysis on observations in B and I bands.
Observational studies of the separations of arm components usually
focus on the SF-arm and the P-arm, observing in (for example) B and K bands.
\citet{beckman90} observed in the B and I bands, finding an offset in one galaxy,
but too flocculent a structure in another.
\citet{grosbol98} measured radial variations in the location of SF-arms and P-arms,
but conclude that uncertainties in the absolute phase offset between them
prevent corotation from being located.  \citet{seigar98} also examined
arms in the K and B bands, observing that the SF-arms appear on the convex
side of the P-arms as expected, but that the P-arms are more tightly wound
than the SF-arms, which is the opposite of the expected result.

All of these investigations are based on observations of the SF-arms.  This
will always be inherently difficult, since the existence, location and shape of the SF-arm are neither
expected, nor observed, to be very regular.  As discussed
in the introduction,
galaxies with regular structure in P-arms will often be classified as flocculent
in optical observations, meaning that they do not possess regular SF-arms.
As pointed out by \citet{elmegreen79}, one would expect the stimulation of
star formation in shock fronts to be followed by a `shuffling' of material,
including feedback from star formation, random structure from the condensation
of dense clouds and so forth.  The appearance of any SF-arm would therefore be
much more irregular, and in the right circumstances will appear entirely
flocculent.  The observation of these arms is further hampered by
the effects of dust obscuration, which can make it difficult to locate their
full extent.  Furthermore, the unknown quantity of the `onset time' for star
formation may in reality be a range of times, leading to star formation occurring
at a range of offsets from the shock.  The prediction of the large scale shock-induced
star formation scenario would be, therefore, that regular P-arms should
produce regular D-arms, and any subsequent SF-arms will be irregular and
poorly localised.

Observations of the P-arms and D-arms, and the offset $\Theta$ between them, avoid
all such issues and would be expected, in galaxies with a regular spiral potential,
to result in a much clearer result.  In the next section, the way in
which corotation might be constrained from such measurements is described.

\subsection{Constraining corotation from the shock--potential offset}

As has already been seen in figure \ref{CMoffsets_m2},
there is a general trend for $\Theta$ to vary in two-armed spirals in a specific
way.  At small radii, $\Theta$ approaches zero, but as the radius approaches corotation,
$\Theta$ moves toward $-\pi$.  Even with $m=4$, although $\Theta$ does not go to
zero at small radii, it still moves toward $-\pi$ with increasing $R$.  The extrapolation
of the curve of $\Theta$, in all these cases, roughly indicates the location of corotation.

\begin{figure}
\includegraphics[width=0.45\textwidth]{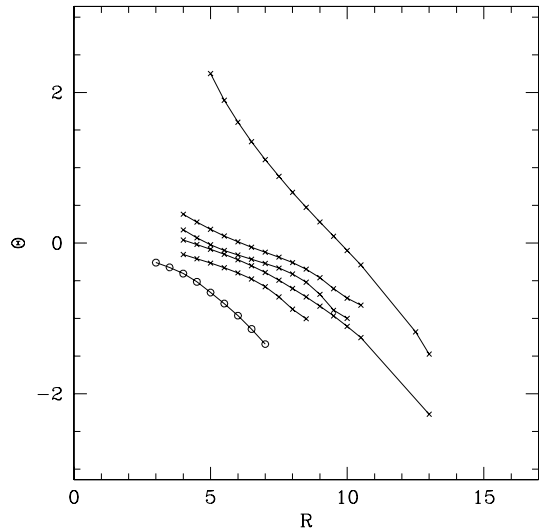}
\caption
{Plots of $\Theta(R)$ with various parameters.  Lines marked with crosses
correspond to (bottom to top) $F=0.03$, standard model, $i=0.15$,
$a = 5$ km s$^{-1}$, $m=4$.  Line with circles corresponds to
$\OmegaP = 19.5$ km s$^{-1}$ kpc$^{-1}$.}
\label{offsets_1}
\end{figure}

The form of $\Theta(R)$ does, of course, depend on the parameters of the spiral potential.
However, the extrapolation of $\Theta$ to find corotation stays approximately unchanged
under all such variations.  Figure \ref{offsets_1} shows the curve for the standard model
compared to the alterations $i=0.15$, $F=0.03$, $a = 5$ km s$^{-1}$, $m=4$ and
$\OmegaP = 19.5$ km s$^{-1}$ kpc$^{-1}$.  Corotation is located at 17 kpc in all of these
models except the last, in which it is at $11.3$ kpc.  The shapes of all the curves indicate
corotation to be in the range 15--20 kpc, except the curve at higher $\OmegaP$ which correctly
indicates a smaller radius of corotation.

\begin{figure}
\includegraphics[width=0.45\textwidth]{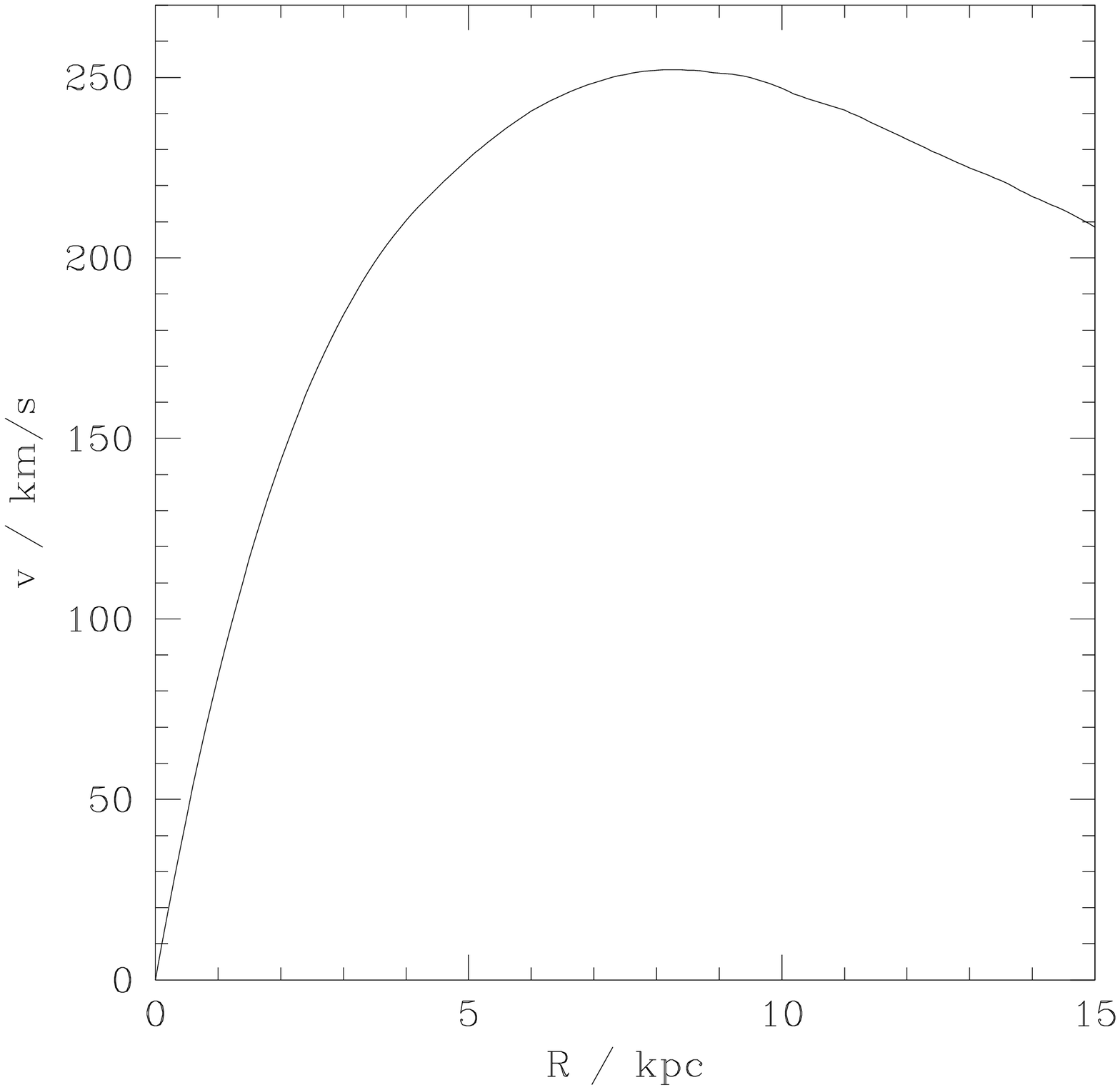}
\caption
{Velocity curve used in \citet{shu73}.}
\label{shuvel}
\end{figure}

\begin{figure}
\includegraphics[width=0.45\textwidth]{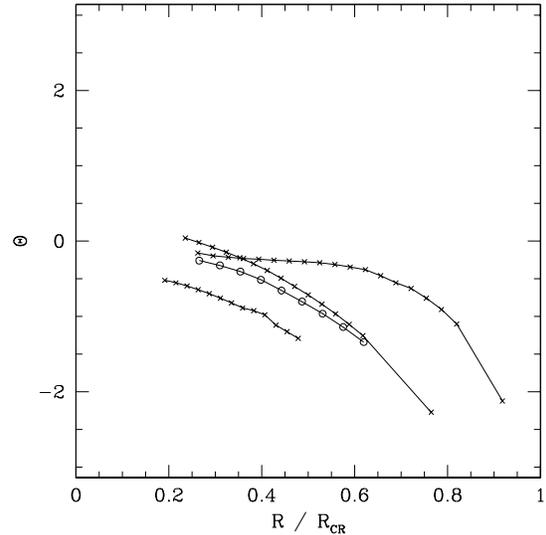}
\caption
{Plots of $\Theta(R)$, plotted against radius relative to corotation, for various parameters.
Lines marked with crosses correspond to (left to right) $\epsilon\sub{d}^{-1} = 10$ kpc,
standard model, velocity curve from figure \ref{shuvel}.
Line marked with circles corresponds to $\OmegaP = 19.5$ km s$^{-1}$ kpc$^{-1}$.}
\label{offsets_2}
\end{figure}

Variations in the velocity curve will also affect the shape of $\Theta(R)$.
In figure \ref{offsets_2}, the resulting plots are compared using a flat
velocity curve, a rising velocity curve in which $\epsilon\sub{d}^{-1} = 10$ kpc
(rather than $1.5$ kpc), and the velocity curve used in \citet{shu73}, which 
is peaked at around 8 kpc, and shown in figure \ref{shuvel}.
The values corresponding to $\OmegaP = 19.5$ km s$^{-1}$ kpc$^{-1}$
are also shown.  Since each of these velocity curves places corotation at a different
radius, the values are plotted against $R / R\sub{corotation}$ in this figure.
Again, in all cases the approximate position of corotation is indicated
by the extrapolation of the curve.

\begin{figure}
\includegraphics[width=0.45\textwidth]{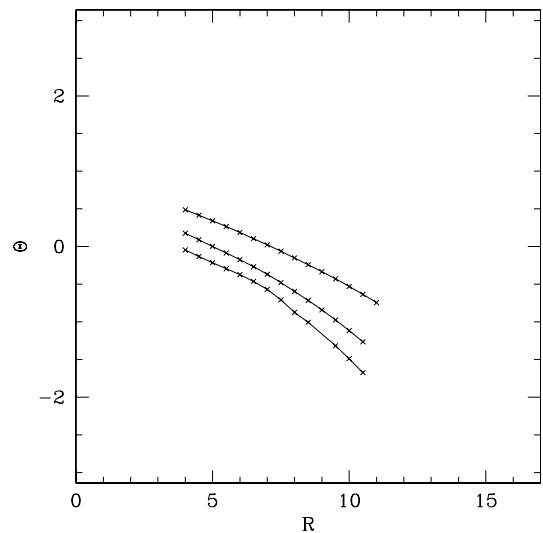}
\caption
{Plots of $\Theta(R)$ for models using a constant $F$ of (bottom to top)
$0.03$, $0.05$ and $0.1$.}
\label{offsets_3}
\end{figure}

Probably the most difficult parameter to constrain in a spiral galaxy is the
relative strength $F$ of the spiral perturbations.  In the standard model,
this takes the value 5\% at $R=8.5$ kpc, and varies as $R \exp(-\epsilon\sub{s} R)$.
In figure \ref{offsets_3}, the shock locations obtained using a constant strength
$F(R)$, at values 3\%, 5\% and 10\%, are shown.  This figure indicates
that the spiral strength introduces the greatest uncertainty into the location
of corotation.  Generally, weaker spirals will form shocks further upstream at smaller
radii than stronger spirals.
If the strength of the spiral perturbation varied systematically across a galactic disc,
for example increasing from 3\% to 10\% (or vice versa), the extrapolation
of $\Theta$ could indicate an inaccurate value for corotation.

\subsection{Example results and accuracy}

To assess the approximate accuracy of the method, we consider in this section
how the semi-analytical results in figures \ref{offsets_1}, \ref{offsets_2} and \ref{offsets_3}
may be used, in conjunction with observational data, to constrain the location of corotation
in spiral galaxies.
We therefore begin with 10 sets of data $\Theta(R)$ for $m=2$.
The general idea is to extrapolate
the downward-curving trend to the point $\Theta = -\pi$, which forms the estimate for corotation.

The radial range over which we have managed to obtain semi-analytic solutions
depends on the model parameters (e.g.\ arm strength, rotation
curve), so that these parameters
affect the range in $\Theta(R)$ for which we have solutions in each case.
Estimates will be more accurate where a larger $\Delta\Theta$ is
available. We wish to assess the approximate accuracy of this method as a function of $\Delta\Theta$.

To this end, we analysed the results at values of $\Delta\Theta$ in steps of $\pi / 8$ up to the
largest value covered by the data.  For each value of $\Delta\Theta$, only those curves were
included whose results cover at least that range, and only the points (at small radii) within
that range were used.  Thus, at small $\Delta\Theta$, all the curves can be extrapolated, but only
the first part of each will be used.  At large $\Delta\Theta$, only a few of the curves have
sufficient data to produce a result, but they will be more accurate.

\begin{figure}
\includegraphics[width=0.45\textwidth]{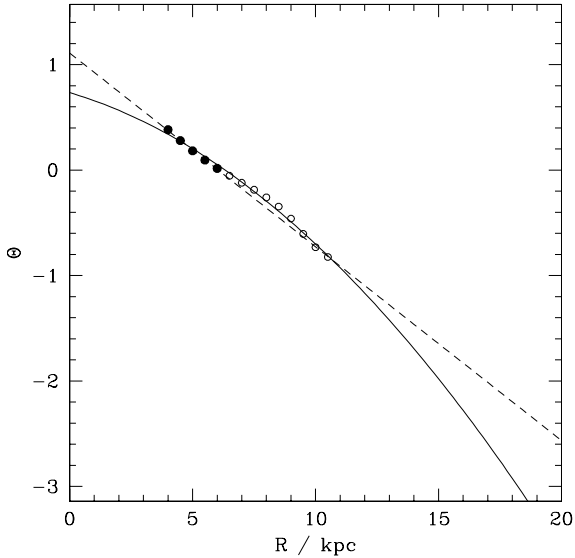}
\caption{The extrapolation of the results with $a=5$ km s$^{-1}$ at two values of $\Delta\Theta$.
At $\Delta\Theta = \pi/8$, only the first five points constitute the data (solid circles).  The
resulting quadratic fit has the wrong curvature, so a linear fit is used (dashed line).  At
$\Delta\Theta = \pi/2$, all the data are included, and the quadratic fit estimates the corotation
radius at $18.6$ kpc.}
\label{cs5fits}
\end{figure}

Any set of data clearly needs to show a decrease in $\Theta$ with $R$ as a minimum requirement
for this method to be applied.  The simplest extrapolation is to apply a least-squares quadratic
fit to the points.  In some cases, the data curve toward positive $\Theta$ over some sections,
in which case a quadratic fit does not extrapolate to $\Theta = -\pi$.  To avoid this problem,
if the coefficient of $R^2$ in the quadratic fit is positive, we use a linear fit instead.
Once a fit has been chosen, the corotation radius is estimated as the intersection of the fit
with $\Theta = -\pi$.  Figure \ref{cs5fits} illustrates some fits for the data with $a=5$ km s$^{-1}$.

\begin{figure}
\includegraphics[width=0.45\textwidth]{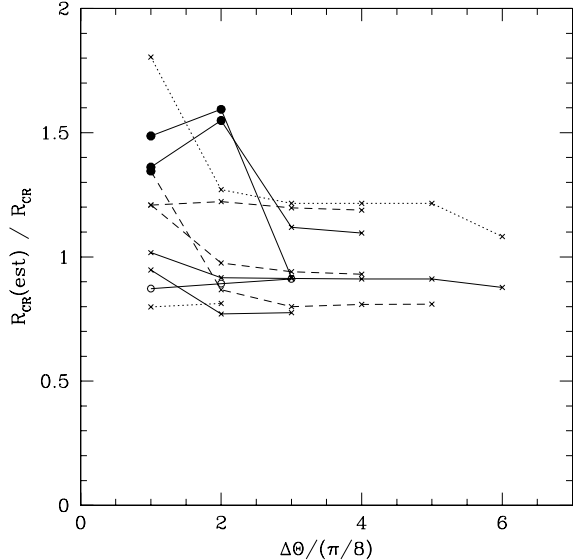}
\caption{The ratio of predicted to actual corotation radius for all the data sets, as a function
of $\Delta\Theta$.  Solid lines with crosses correspond to (bottom to top at left end)
$F=0.03$, standard model, $a=5$ km s$^{-1}$, $i=0.15$.  The solid line with circles
corresponds to $\OmegaP = 19.5$ km s$^{-1}$ kpc$^{-1}$.
Dashed lines are models with constant
$F$ with values (bottom to top at right end) 3\%, 5\%, 10\%.  Dotted lines are alternative
velocity curves (lower: $\epsilon\sub{d}^{-1}=10$ kpc, upper: figure \ref{shuvel}).
Solid circles indicate results where a linear (rather than quadratic) extrapolation was used.
}
\label{corotationresults}
\end{figure}

The scatter in the resulting predicted values for corotation over the 10 sets of data
gives an indication of the accuracy of this method.  The results are plotted in figure
\ref{corotationresults} for all values of $\Delta\Theta$.  At an optimal value of
$\Delta\Theta \simeq 3\pi/8$ to $4\pi/8$, most of the data sets cover a sufficient range
to be included, and the spread is some 25\%.  We estimate, therefore, that this method
should be accurate to around 25\% if a sufficiently large $\Delta\Theta$ is available
in the observed data, and that $\Delta\Theta \ga \pi/4$ is a reasonable guide to this
requirement.

\section{Conclusions}
\label{conclusions}

In this paper, we re-examine the response of an isothermal gas disc to a model spiral
potential, leading to a large-scale flow containing a single shock.  The established
semi-analytical method was compared to results from numerical techniques, with good agreement
within the limitations of each method.  We found that:

\begin{enumerate}
\item{Results from two-dimensional SPH fail to resolve
the shock sufficiently sharply to determine its location.  More work on the details of
the SPH implementation may improve results.}
\item{Much better results were obtained with the PPMLR code \CMHOG.}
\item{In parameter regimes where a semi-analytical solution cannot be found, in general the numerical
results failed to show a simple single-shock solution.}
\end{enumerate}

We discuss the appearance of spiral arms in galaxies, and the existence of three separate
arms, the P-arm (in the stellar disc, approximately tracing the gravitational
Potential), the D-arm (the shock as traced by Dust) and the SF-arm
(an arm traced by enhanced Star Formation).  Even in galaxies classified as optically
flocculent, a regular structure in the P-arms and D-arms should be common.
The angular offset $\Theta(R)$ between the
D-arm and P-arm is proposed as an observable function, avoiding all the difficulties
inherent in tracing the shape and location of the SF-arm.

Finally, we show how this function could be extrapolated to provide an estimate
of the corotation radius, and assess the accuracy of the method as a function of
the range $\Delta\Theta$ available in the observations.
We estimate that given $\Delta\Theta \ga \pi/4$, the method is accurate to around 25\%.


\section*{Acknowledgements}

Thanks are due to Peter Teuben for providing the \CMHOG\ code.  Advice and comments
are acknowledged from Jim Pringle, Ron Allen, Steve Lubow, Jerry Sellwood,
Eve Ostriker and Jim Stone.  Thanks are also due to Jonathan Gair for suggesting
the change of variables to avoid the singularity in the semi-analytical equations.
Finally, we thank the referee for useful comments, and for suggesting the addition
of a quantitative assessment of the accuracy of the method.



\appendix
\section{Method for locating semi-analytical solutions}
\label{appendix}

In this appendix, the method for locating solutions in the semi-analytical approach
is described for convenience.  The method is almost the same as that described in \citet{shu72}
and \citet{roberts69}.

The differential equations describing the response of the gas,
along a streamline with average radius $R_0$, are as follows:
\[
	\frac{\partial u_{\eta 1}}{\partial \eta} = U(u_{\eta 0} + u_{\eta 1})
	\frac{ 2 u_{\xi 1} - F R_0 \Omega \sin \eta }{ (u_{\eta 0} + u_{\eta 1})^2 - a^2}
\]
\[
	\frac{\partial u_{\xi 1}}{\partial \eta} = -V \frac{u_{\eta 1}}{u_{\eta 0} + u_{\eta 1}}
\]
where
\[
	u_{\eta 0} = R_0 (\Omega - \OmegaP) \sin i
\]
\[
	u_{\xi 0} = R_0 (\Omega - \OmegaP) \cos i
\]
\[
	U = \frac{\sin i}{m} R_0 \Omega
\]
\[
	V = \frac{\sin i}{m} \frac{\rm d}{{\rm d}R}(R^2 \Omega)
	= \frac{\sin i}{m} \frac{R_0 \kappa^2}{2 \Omega}.
\]

The resulting gas density $\sigma = \sigma_0 + \sigma_1$
relative to the unperturbed density $\sigma_0(R)$ is given by
\[
	(\sigma_0 + \sigma_1)(u_{\eta 0} + u_{\eta 1}) = \sigma_0 u_{\eta 0}.
\]

It is convenient to transform these equations into dimensionless form with $(u_{\eta 1},u_{\xi 1})$
replaced by the dimensionless $(u,v)$ defined by
\[
	u = \frac{u_{\eta 1}}{\sqrt{2UV}}
\]
\[
	v = \frac{u_{\xi 1}}{V}
\]
by first defining the dimensionless parameters
\[
	f = \frac{F R_0 \Omega}{2V}
\]
\[
	\nu = \frac{-u_{\eta 0}}{\sqrt{2UV}}
\]
\[
	x = \frac{a^2}{2UV}
\]
so that the final dimensionless equations to be solved are as follows:
\begin{equation}
	\frac{\partial u}{\partial \eta} = (u-\nu) \frac{v - f \sin \eta}{(u-\nu)^2 - x}
\label{floweq1}
\end{equation}
\begin{equation}
	\frac{\partial v}{\partial \eta} = \frac{u}{\nu-u}
\label{floweq2}
\end{equation}

For low spiral strength $F$ or sufficently high or low sound speed $a$, solutions
will be entirely supersonic or entirely subsonic, and will not contain a shock.
For parameters where $u_{\eta}$ crosses the value $a$, a shock must exist.  The method
for locating such a solution containing a single shock (if it exists) is as follows.

As can be immediately seen from equations \ref{floweq1} and \ref{floweq2},
singularities occur at three points: $u=\nu$, $u=\nu - \sqrt{x}$ and $u=\nu + \sqrt{x}$.
These correspond, respectively, to $u_\eta = 0$, $u_\eta = -a$ and $u_\eta = a$.  For
a solution flowing through the shock, the first two conditions should never occur, since
gas flow through a trailing spiral (inside corotation) should always be moving outward
in $\eta$.
The third singularity corresponds to the sonic point, at which the gas undergoes the transition
to supersonic flow after the shock (the gas returns to subsonic flow in the
shock itself).  The value of $\eta$ at which this sonic point occurs is denoted
$\eta\sub{SP}$.
The singularity at this point is removed as long as $v(\eta\sub{SP})=f\sin\eta\sub{SP}$.
This requirement fixes the boundary conditions;
given any guessed value for $\eta\sub{SP}$, the starting values of $u$ and $v$ at
this point must be $u=\nu + \sqrt{x}$ and $v=f\sin\eta\sub{SP}$.

Solutions are therefore calculated starting at $\eta\sub{SP}$, and proceeding forward in $\eta$
(the subsonic branch) and backward in $\eta$ (the supersonic branch).  The determination
of the shock location and the procedure for adjusting $\eta\sub{SP}$ to find a solution
are described below.

The presence of the singularity at $\eta\sub{SP}$ can create difficulties for numerical integration.
\citet{shu73} used a series expansion in $\eta$ around $\eta\sub{SP}$ to overcome this
problem.  An alternative is to rewrite the equations in new variables,
which hide the singularity, as follows (J. Gair, private communication).
Start by defining
\[
	u = \nu + \sqrt{x} + \tilde{u}
\]
\[
	v = f \sin\eta + \tilde{v}
\]
and then define
\[
	\tilde{w} = \tilde{u}^2.
\]
The differential equations now become
\[
	\frac{\partial \tilde{w}}{\partial \eta} = \frac{2 \left( \sqrt{x}\pm\sqrt{\tilde{w}} \right) \tilde{v}}
		{2\sqrt{x}\pm\sqrt{\tilde{w}}}
\]
\[
	\frac{\partial \tilde{v}}{\partial \eta} = -1 -f\cos\eta -\frac{\nu}{\sqrt{x}\pm\sqrt{\tilde{w}}}
\]
where the positive root of $\tilde{w}$ is taken where $\tilde{u}>0$ (i.e.\ along the supersonic branch)
and the negative root is taken where $\tilde{u}<0$ (the subsonic branch).  In this form,
the equations can be integrated along each branch separately, starting at $\eta\sub{SP}$,
and $\tilde{w}$ and $\tilde{v}$ transformed back to find $u$ and $v$.  The singularity
is hidden and replaced by a forbidden region of phase space, namely the requirement
that $\tilde{w}>0$.  This will be satisfied so long as the correct boundary
condition is used ($\tilde{v}(\eta\sub{SP})=\tilde{w}(\eta\sub{SP})=0$).

The procedure for locating the shock is as follows.  The flow is integrated from $\eta\sub{SP}$
along the supersonic and subsonic branches, to give $u\sub{sup}(\eta)$ and $u\sub{sub}(\eta)$.
Generally, the supersonic branch will eventually return to the singularity at $u=a$ and cannot
be further integrated, while the subsonic branch either tends asymptotically to $u=0$, or
also reaches $u=a$.
If a consistent solution exists, then the shock will terminate both branches before they
reach singularities.

\begin{figure}
\includegraphics[width=0.45\textwidth]{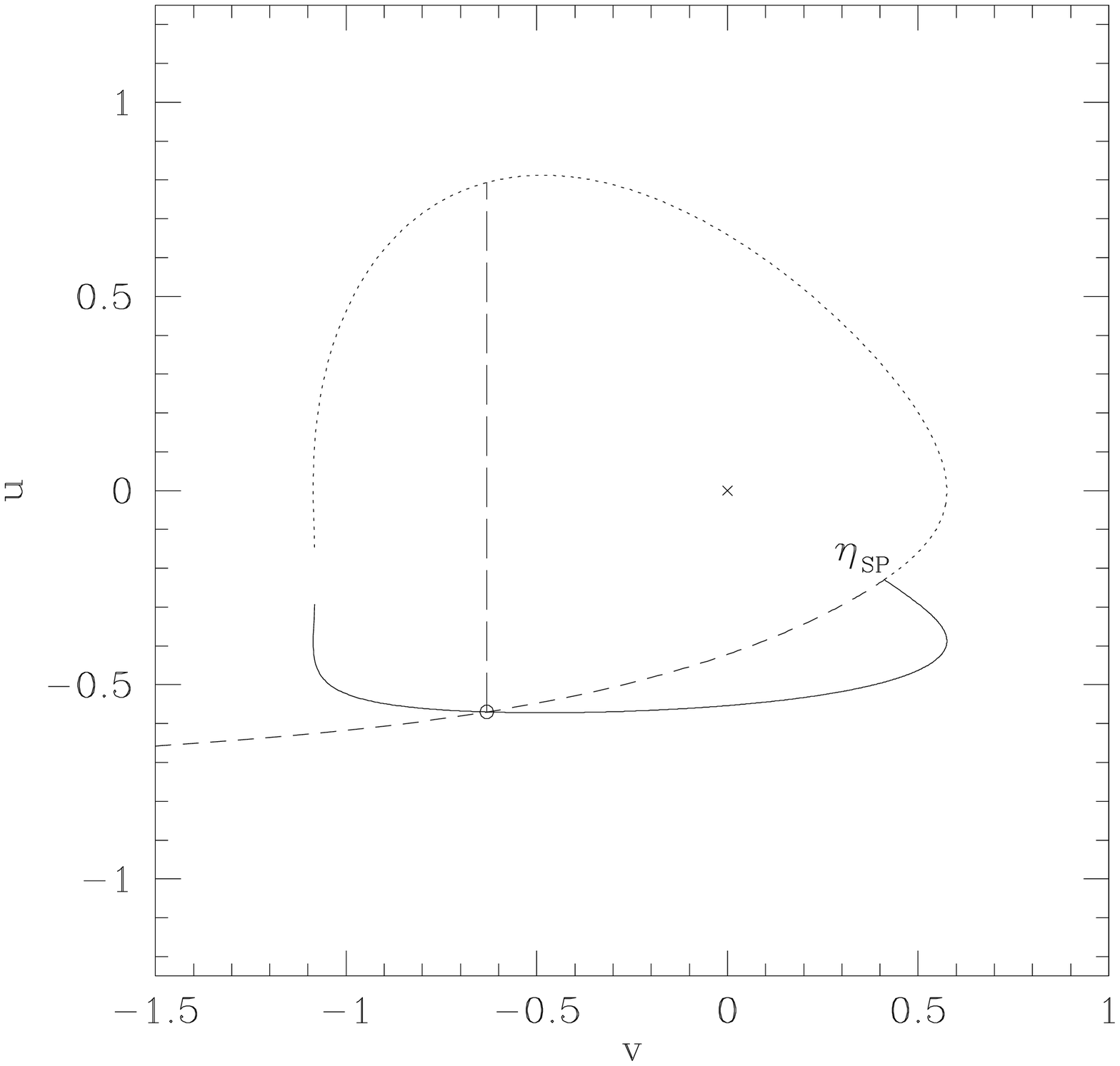}
\caption
{The process of locating the shock.  Integration starts at the sonic point (labelled $\eta\sub{SP}$)
along the supersonic (dotted) and subsonic (short dashed) branches.  The post-shock branch
(solid) is calculated for each point along the supersonic branch using the jump conditions.
It crosses the subsonic branch at the appropriate location for the shock (circle).
The shock, indicated by the long dashed line, closes the velocity curve.  Flow proceeds
in an anticlockwise direction.  The value of $\eta\sub{SP}$ should now be adjusted until
the period in $\eta$ of the solution reaches $2\pi$.
The unperturbed flow corresponds to $u=v=0$ (cross).}
\label{shocklocation}
\end{figure}

For every point along the supersonic branch $u\sub{sup}$, the isothermal jump conditions
can be used to calculate the post-shock velocity $u\sub{shock}$ that the flow would have if the shock were
to occur at that particular value of $\eta$.  Since the shock is perpendicular
to the $\eta$ direction and the gas is isothermal, the jump condition is simply
\[
	u\sub{sup} u\sub{shock} = a^2
\]
and so the post-shock branch is defined (in terms of the dimensionless parameters) by
\[
	u\sub{shock}(\eta) = \nu + \frac{x}{u\sub{sup}(\eta) - \nu}.
\]
Through the shock, $v$ should be continuous, but $u$ should drop from $u\sub{sup}$ to
$u\sub{shock}$.  In a consistent solution, the shock joins the supersonic branch
to the subsonic branch at the same value of $v$.  Therefore, the only consistent
location for the shock exists where the post-shock branch crosses
the subsonic branch in the $(v,u)$ plane.  This process is shown visually in
figure \ref{shocklocation}.  The starting value of $\eta\sub{SP}$ must be adjusted
until a shock location appears, at a point on each branch before they reach a singularity.
For some parameters, no value of $\eta\sub{SP}$ will produce a shock.

The location of the crossing point is
found numerically by minimising $u\sub{sub}(\eta\sub{sub}) - u\sub{shock}(\eta\sub{sup})$
with respect to $\eta\sub{sub}$ and $\eta\sub{sup}$, which are the values of $\eta$
on the subsonic and supersonic branch (respectively) at which the shock occurs.
The period in $\eta$ of the solution
will therefore be $\eta\sub{sup} - \eta\sub{sub}$.  In general, for a given value
of the sonic point $\eta\sub{SP}$, this period will not be $2 \pi$.  The starting
value of $\eta\sub{SP}$ must therefore be adjusted, repeating the shock location procedure
at each value, until a solution is found with a period of $2 \pi$.
If such a value can be found, it represents the only physical solution to the flow
equations, containing one isothermal shock.  The density $\sigma(\eta)$ can
be calculated from the values of $u$.

Solutions in the base-subsonic region are particularly hard to find.
This is because the integrated branches depend very sensitively on the
value of $\eta\sub{SP}$.  In some cases, the range of values for which
a solution containing a shock exists can have a width of less than $\pi / 1000$,
and so can be difficult to locate.  Outside this range, the calculated branches
run into singularities too rapidly.

\end{document}